\documentclass[%
 reprint,
 superscriptaddress,
 preprintnumbers,
 amsmath,amssymb,
 aps,
]{revtex4-2}

\usepackage[utf8]{inputenc}
\usepackage{graphicx}
\usepackage{grffile}
\usepackage[usenames,dvipsnames]{xcolor}
\usepackage{amsmath}
\usepackage[normalem]{ulem}
\usepackage[resetlabels, labeled]{multibib}
\usepackage{appendix}
\newcites{S}{References Supplementary Materials}
\definecolor{orange}{rgb}{1,0.5,0}
\definecolor{goodgreen}{rgb}{0.1,0.5,0}
\definecolor{goodred}{rgb}{0.7,0,0}
\usepackage{lineno}
\usepackage[colorlinks,urlcolor=goodgreen,citecolor=blue,linkcolor=goodred]{hyperref}
\usepackage{float}
\usepackage{babel}
\graphicspath{ {images/} }

\makeatother

\begin{document}

\title{Microwave-assisted thermoelectricity in \texorpdfstring{$S$-$I$-$S'$}{S-I-S'} tunnel junctions}% Force line breaks with \\

\newcommand{\orcid}[1]{\href{https://orcid.org/#1}{\includegraphics[width=8pt]{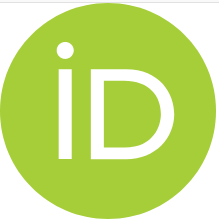}}}

\author{A. Hijano\orcid{0000-0002-3018-4395}}
\email{alberto.hijano@ehu.eus}
\affiliation{Centro de F\'isica de Materiales (CFM-MPC) Centro Mixto CSIC-UPV/EHU, E-20018 Donostia-San Sebasti\'an,  Spain}
\affiliation{Department of Condensed Matter Physics, University of the Basque Country UPV/EHU, 48080 Bilbao, Spain}
\author{F. S. Bergeret\orcid{0000-0001-6007-4878}}
\email{fs.bergeret@csic.es}
\affiliation{Centro de F\'isica de Materiales (CFM-MPC) Centro Mixto CSIC-UPV/EHU, E-20018 Donostia-San Sebasti\'an,  Spain}
\affiliation{Donostia International Physics Center (DIPC), 20018 Donostia--San Sebasti\'an, Spain}
\author{F. Giazotto}
\email{francesco.giazotto@sns.it}
\affiliation{NEST Istituto Nanoscienze-CNR and Scuola Normale Superiore, I-56127 Pisa, Italy}
\author{A. Braggio\orcid{0000-0003-2119-1160}}
\email{alessandro.braggio@nano.cnr.it}
\affiliation{NEST Istituto Nanoscienze-CNR and Scuola Normale Superiore, I-56127 Pisa, Italy}

%\affiliation{%



\date{\today}% It is always \today, today,

\begin{abstract}
Asymmetric superconducting tunnel junctions with gaps $\Delta_1>\Delta_2$ have been proven to show a peculiar nonlinear bipolar thermoelectric effect. This arises due to the spontaneous breaking of electron-hole symmetry in the system, and it is maximized at the matching-peak bias $|V|=V_p=(\Delta_1-\Delta_2)/e$. In this paper, we investigate the interplay of photon-assisted tunneling (PAT) and bipolar thermoelectric generation. In particular, we show how thermoelectricity, at the matching peak, is supported by photon absorption/emission processes at the frequency-shifted sidebands $V=\pm V_p+n\hbar\omega$, $n \in \mathbb{Z}$. This represents a sort of \emph{microwave-assisted thermoelectricity}. We show the existence of multiple stable solutions, being associated with different photon sidebands, when a load is connected to the junction. Finally, we discuss how the nonlinear cooling effects are modified by  the PAT. The proposed device can  detect millimeter wavelength signals by  converting a temperature gradient into a thermoelectric current or voltage.
\end{abstract}

\maketitle

%\tableofcontents

\section{Introduction}\label{introduction} 
Hybrid superconducting systems with explicit broken particle-hole symmetry show unipolar thermoelectricity~\cite{bergeret2018colloquium,heikkila2019thermal,machon2013nonlocal,ozaeta2014predicted,kolenda2016observation,Heikkila:2018,Chakraborty:2018,strambini2022superconducting,Geng:2022}.  The particle-hole symmetry around the Fermi surface of Bardeen-Cooper-Schrieffer (BCS) superconductors can be broken, for instance, in superconductor/ferromagnet hybrid structures. The magnetic proximity effect in a thin superconductor-ferromagnetic insulator bilayer causes an almost homogeneous spin splitting of the density of states (DOS)~\cite{Meservey:1994}. If the electronic transport is spin-polarized, for example via a tunneling spin-filter, ~\cite{Moodera:2007,Miao:2014,DeSimoni:2018,Morten:2005}, the DOS contribution of one spin component becomes predominant over the other one, leading to an effective  particle-hole symmetry breaking~\cite{ozaeta2014predicted}.

It has recently been theoretically~\cite{Marchegiani:2020a,Marchegiani:2020b,Marchegiani:2020c} and experimentally~\cite{Germanese:2022,Germanese:2022b} shown that superconducting tunnel junctions, where the Josephson coupling is properly suppressed, develop a large thermoelectric effect if the electrode with the larger gap has a higher temperature. In contrast to systems with magnetic proximity effect, in superconducting tunnel junctions the electron-hole symmetry is broken by the combination of a sufficiently strong thermal gradient and a monotonously decreasing DOS which induces spontaneous voltage polarization. The resulting thermoelectricity is \emph{bipolar} and strongly nonlinear.

We focus here on photon-assisted tunneling (PAT) which has been extensively studied in the dissipative regime~\cite{Tien:1963,Sweet:1970,Hamilton:1970,Kofoed:1974,Tucker:1979,Tucker:1985,Hamasaki:1982,Kashinje:1986,Sun:1999,Leone:2001,Kot:2020}. However the  influence of PAT on the recently observed bipolar thermoelectricity is still unexplored. 
In this work, we address microwave-assisted thermoelectricity by investigating the interplay between PAT and bipolar thermoelectricity in a superconductor-insulator-superconductor ($S$-$I$-$S'$) tunnel junction. This kind of tunnel junctions normally work at very low temperatures in order to preserve the superconductivity. It is worth noting that thermoelectricity at these temperatures is very strong~\cite{Germanese:2022} due to the nonlinearity of the effect. The obtained results show intriguing perspectives for application as millimeter wavelength signals detectors~\cite{Chakraborty:2018}, flux controlled high-frequency oscillators~\cite{Marchegiani:2020c}, controlled generators~\cite{Marchegiani:2020b}, superconducting qubits~\cite{Houzet:2019,Pan:2022,Liu:2022-arXiv,Diamond:2022,Marchegiani:2022}, quantum sensors~\cite{Paolucci:2023-arXiv,Germanese:2023-Patent} and, more broadly, in the emergent field of superconducting quantum technologies.

This paper is organized as follows: In Sec.~\ref{model} we present the basic equations describing the DC and AC tunneling charge and heat currents through a tunnel junction. In Sec.~\ref{Sec:results} we numerically obtain the $I(V)$ characteristic curves and study the thermoelectric power output. Finally, we study the impact of the AC voltage source on the cooling power of the junction. We summarize the results in Sec.~\ref{conclusions}.

%
%
%%%%%%%%%%%%
\begin{figure*}[!t]
\centering
  \includegraphics[width=0.8\textwidth]{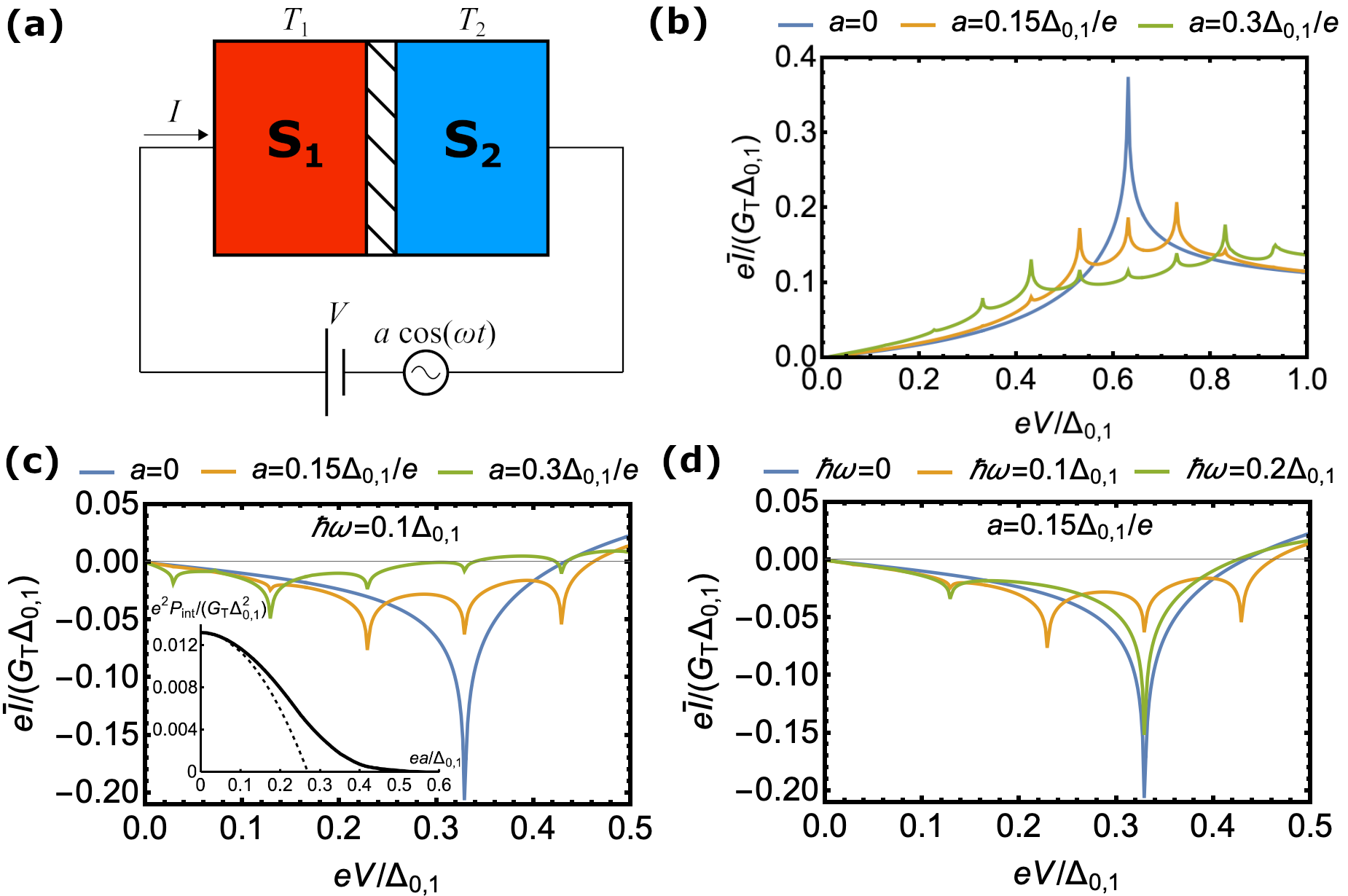}
\caption{(a) Simple circuit scheme of the photon-assisted bipolar thermoelectricity. The two superconductors have a different gap $\Delta_1>\Delta_2$ and are subject to a temperature difference $T_1>T_2$. The $S$-$I$-$S'$ junction is powered by a DC and an AC voltage source. (b) Equilibrium $\bar{I}(V)$ curves for a finite AC voltage with $\hbar\omega=0.1\Delta_{0,1}$ and different amplitudes $a$. Time averaged quasiparticle current $\bar{I}(V)$ dependence on the amplitude (c) and angular frequency (d) of the AC field. The inset in panel (c) shows the integrated thermoelectric power as a function of $a$ (solid line) and the predicted behaviour by the small signal limit (dashed line). The temperatures are $T_1=T_2=0.4T_{c,1}$ in (b), and $T_1=0.7T_{c,1}$ and $T_2=0$ for panels (c) and (d).}\label{Fig:I_vs_V}
\end{figure*}
%%%%%%%%%%%
%
%
\section{The Model}\label{model}
We consider  an $S$-$I$-$S'$ tunnel junction where each superconducting electrode is kept in thermal equilibrium, see Fig.~\ref{Fig:I_vs_V}(a). The gap of the left superconductor ($S_1$) is considered to be higher than that of the right superconductor ($S_2$), $\Delta_1>\Delta_2$. This can be done by using different superconducting materials for each electrode or by taking advantage of the inverse proximity effect by attaching a normal layer to the right superconductor, which effectively suppresses the gap~\cite{Germanese:2022}. The asymmetry parameter $r=\Delta_{0,2}/\Delta_{0,1}<1$ quantifies the asymmetry between the two terminals, where $\Delta_{0,\alpha}$ is the zero-temperature gap of electrode $\alpha$. 
If the electrodes are powered by a constant voltage source $V$, the chemical potential of the electrodes are shifted by $\mu_1-\mu_2=-eV$. 

The tunneling current between the two superconductors has three contributions: the quasiparticle current, the Cooper pair current, and the interference current which gives the interference contribution associated with breaking and recombination processes of Cooper pairs in different electrodes~\cite{Barone_Paterno,Josephson:1962,Harris:1974,Gulevich:2017}. The latter two contributions stem from the Josephson coupling, and they depend on the phase difference between the superconductors. 
At finite bias, those terms oscillate between positive and negative values, and might be detrimental to a stable thermoelectric effect~\cite{Marchegiani:2020c}. Therefore, we assume that the Josephson coupling is sufficiently weak~\cite{Barone:1975,Giazotto:2006,Fornieri:2017,Germanese:2022}, and neglect those terms such that we consider only the quasiparticle (qp) current which is phase-independent. The Josephson current can also be suppressed by applying a suitable in-plane magnetic field or by applying a small out-of-plane magnetic field in a superconducting quantum interference device (SQUID) as in \cite{Germanese:2022,Germanese:2022b}.

The DC tunneling qp charge $I_\alpha$ and heat $\dot{Q}_\alpha$ currents flowing out from electrode $\alpha=1,2$ are given by
\begin{multline}\label{Eq:current}
    \begin{pmatrix}
    I_\alpha\\
    \dot{Q}_\alpha
    \end{pmatrix}=
    \frac{G_T}{e^2}\int \mathrm{d}E 
    \begin{pmatrix}
    -e\\
    E-\mu_\alpha
    \end{pmatrix}
    N_\alpha(E-\mu_\alpha)N_{\bar{\alpha}}(E-\mu_{\bar{\alpha}})\\
    (f_\alpha(E-\mu_\alpha)-f_{\bar{\alpha}}(E-\mu_{\bar{\alpha}}))\; ,
\end{multline}
with $\bar{\alpha}$ the opposite side with respect to $\alpha$.
Here, $-e$ is the electron charge, $G_T$ is the conductance of the junction, and $f_\alpha(E)=[\exp{(E/k_B T_\alpha)}+1]^{-1}$ is the $\alpha$-lead Fermi-Dirac distribution. The BCS DOS is given by $N_\alpha(E)=\mathrm{Re}\{-i(E+i\Gamma_\alpha)/\sqrt{\Delta_\alpha^2-(E+i\Gamma_\alpha)^2}\}$, where $\Gamma_\alpha$ is the Dynes parameter~\cite{Dynes:1978}, describing inelastic processes. The Dynes parameter accounts for the broadening of the BCS coherent peaks at $E=\pm\Delta$ in the DOS. In the calculations below we set $\Gamma_1=\Gamma_2=10^{-4}\Delta_{0,1}$, which is a reasonable value typically found in experiments with high-quality tunnel junctions~\cite{Giazotto:2004,Timofeev:2009,MartinezPerez:2015,Julin:2016}.

From Eq.~\eqref{Eq:current}, one can easily check  that the $S$-$I$-$S'$ is a reciprocal electric device. Indeed, the DOS of the superconducting electrodes is electron-hole symmetric, so that the quasiparticle charge current in the junction is odd in voltage $I(-V)=-I(V)$ where we focus here on the current flowing out of the left electrode, $I \equiv I_1$. Moreover, due to energy conservation, $\dot{Q}_1+\dot{Q}_2-\dot{W}=0$, where $\dot{W}=-V I$ describes the electric power generated (dissipated) for $\dot{W}>0$ ($\dot{W}<0$) by the junction. Here we use the active sign convention by  considering the electrical work done by the junction over its surroundings as positive. At the equilibrium $T_1=T_2$ the junction is dissipative ($IV>0$). When the temperature of the left electrode, with a bigger gap, is higher than that on the right electrode, $T_1\gtrsim T_2/r$, it is possible to generate bipolar thermoelectricity~\cite{Marchegiani:2020a,Germanese:2022} with the current $I$ flowing against the bias ($IV<0$). This occurs  at subgap voltages $e|V|\lesssim\Delta_1+\Delta_2$~\cite{Marchegiani:2020a,Marchegiani:2020b,Marchegiani:2020c}. Furthermore, from the negative differential conductance at $V\approx 0$, the thermocurrent grows to a maximum around the matching peak $eV_p=\Delta_1-\Delta_2$ and for $e|V|\gtrsim\Delta_1+\Delta_2$ it becomes again dissipative. In an open circuit configuration, a thermoelectric Seebeck voltage $\pm V_S$ is induced, for which $I(\pm V_S)=0$. The double sign of the Seebeck voltage reflects the \emph{bipolarity} of the thermoelectric effect, which is a consequence of the reciprocity of the device.

If an AC source is included in addition to the DC voltage, $V(t)=V+a\cos{(\omega t)}$, for instance by placing the junction in a microwave field, the average current is not anymore simply given by Eq. \eqref{Eq:current}. The time-averaged DC tunnel (charge$|$heat) currents $\bar{I}|\bar{\dot{Q}}=(1/T)\int_{0}^{T}\!\!\mathrm{d}t\, I|\dot{Q}(t)$ take the form~\cite{Tien:1963,Falci:1991,Kot:2020}
\begin{equation}\label{Eq:current_ac}
    \begin{pmatrix}
    \bar{I}\\
    \bar{\dot{Q}}
    \end{pmatrix}=
    \sum_{n=-\infty}^{\infty}J_n^2\left(\frac{ea}{\hbar\omega}\right)
    \begin{pmatrix}
    I\left(V-n\hbar\omega/e\right)\\
    \dot{Q}\left(V-n\hbar\omega/e\right)
    \end{pmatrix}\; ,
\end{equation}
where $T=2\pi/\omega$ is the period of the AC voltage. This is the standard result for PAT where $J_n(x)$ is the $n$-th Bessel function of the first kind. In Eq. \eqref{Eq:current_ac} we assume that the frequency of the AC voltage is small enough so that it modulates the potential energy of the quasiparticles adiabatically~\cite{Tucker:1985,Wingreen:1993,Jauho:1994}. In this approximation, the AC frequency is necessarily bounded by the plasma frequency of the two electrodes and the driving frequency needs to be $\hbar\omega < 2\Delta$, in order to neglect high-order processes in the current due to the direct breaking of Cooper pairs due to photon absorption. At the same time, we will mainly focus on quite small amplitudes of the voltage oscillations $ea\ll \Delta$, as we will explain in Sec.~\ref{Sec:results}, since high amplitudes are detrimental for the thermoelectric effect restoring the usual dissipative behaviour at high energies. Finally, the averaged current $\bar{I}$ is also reciprocal $\bar{I}(-V)=-\bar{I}(V)$ due to the reciprocity of the junction $I(V)$. For this reason, in the following, we show only results for positive biases.

\section{Results}\label{Sec:results}
In Fig.~\ref{Fig:I_vs_V}(b) we show the $\bar{I}(V)$ characteristic curve for the equilibrium case $T_1=T_2=0.4T_{c,1}$ in the presence of PAT with $\hbar\omega=0.1\Delta_{0,1}$ for different amplitudes. As expected for equilibrium temperatures, the junction has only a dissipative behaviour and the PAT introduces different sidebands nearby the matching peak $V_p$. Indeed, if the frequency $\hbar\omega/e$ of the AC voltage is larger than the width of the matching peak and the amplitude $a$ is not too high (see later) we expect to see a weighted replication of the DC characteristics displaced in voltage in the sidebands. According to Eq. \eqref{Eq:current_ac} one easily finds the position of the matching peak sidebands at $eV=eV_p+n\hbar\omega$, $n \in \mathbb{Z}$.

In the bottom panels (c-d) of Fig.~\ref{Fig:I_vs_V} we show the $\bar{I}(V)$ characteristics when the junction is subject to a temperature difference $T_1=0.7T_{c,1}$ and $T_2=0$, with $T_{c,\alpha}$ the critical temperatures of the electrodes. The temperature difference of the electrodes has been chosen in order to maximize the thermoelectric effect, as previously stated~\cite{Marchegiani:2020a}. Figures~\ref{Fig:I_vs_V}(c-d) show the effect of $a$ and $\omega$ on the $\bar{I}(V)$, respectively. For $a=0$ or $\omega=0$ (blue line), the voltage source has no AC component, so the electric current displays the conventional behaviour for a single thermoelectric peak centered at $eV_p=\Delta_1-\Delta_2$. The sharpness of the thermoelectric peak depends on the phenomenological Dynes parameter, with lower $\Gamma$ favoring sharper peaks. As expected from Eq. \eqref{Eq:current_ac}, for a finite AC voltage source the qp current presents multiple thermoelectric peaks which are separated periodically in voltage. As the value of $a$ increases, the height of the main thermoelectric peak decreases in absolute value while the sidebands increase in size. Clearly, at fixed $\omega$, by changing the amplitude $a$ the weights given by the Bessel functions change such that the height of the sideband peaks change correspondingly. Physically, for $V>0$, the sideband thermoelectric peaks in the averaged current correspond to processes where photons are absorbed (emitted) respectively for $n < 0$ ($n > 0$) to support thermoelectricity at the sideband voltages.
Note that the signature of the thermoelectric peak corresponding to the sidebands can also be found in the dissipative regime as a dip of the $\bar{I}(V)$ characteristics [see the dip at $eV/\Delta_{0,1}\approx 0.45$ of the green line in Fig.~\ref{Fig:I_vs_V}(c)]. All sidebands do not necessarily become thermoelectric, being weighted by a different Bessel function and mixing different channels in the thermoelectric and dissipative regimes. In general, the $I(V)$ characteristic at sufficiently high voltages remains in the dissipative regime.
%
%
%%%%%%%%%%%%
\begin{figure*}[!t]
\centering
  \includegraphics[width=0.8\textwidth]{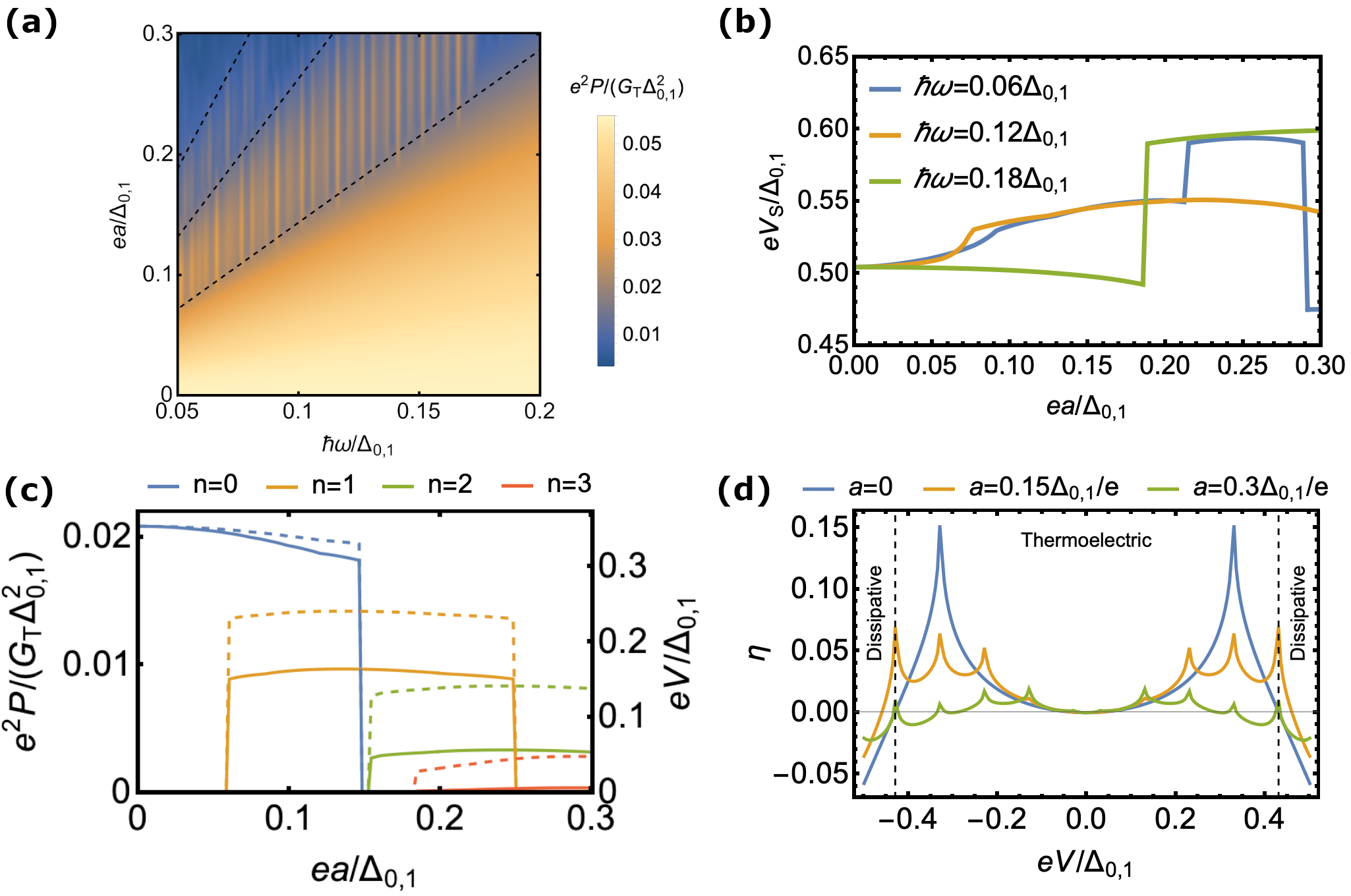}
\caption{(a) Maximum power and (b) Seebeck voltage as a function of the amplitude and angular frequency of the AC field. (c) Power (solid) and voltage (dashed) of the stable points of the $\bar{I}(V)$ characteristic curve for a fixed load $R=6G_T^{-1}$. (d) Efficiency $\eta$ of the thermoelectric effect. In panels (c-d) the frequency is set to $\hbar\omega=0.1\Delta_{0,1}$. The temperatures of the superconducting electrodes are $T_1=0.7T_{c,1}$ and $T_2=0$ for panels (a), (c) and (d), and $T_1=0.6T_{c,1}$ and $T_2=0$ for panel (b).}\label{Fig:power}
\end{figure*}
%%%%%%%%%%%
%
%
In other words, the AC source seems to redistribute the thermoelectric power to different sidebands. This can be better quantified by looking at the integrated thermoelectric power, defined as
\begin{equation}
\label{Pbar}
    P_{\mathrm{int}}=-\int_{0}^{V_S}\mathrm{d}V\  \bar{I}(V)
\end{equation}
which decreases with increasing $a$ [see inset of~\ref{Fig:I_vs_V}(c)], and remains almost constant under variations of $\omega$ (not shown). The dashed line in the inset corresponds to the power given in the small-signal limit. Indeed, when $x=(ea/\hbar\omega)\ll 1$, the variation of the averaged current may be obtained by expanding the Bessel functions to the lowest order $J_0(x)\approx 1-x^2/4$ and $J_{\pm n}(x)\approx (\pm x/2)^n/n!$ in Eq. \eqref{Eq:current_ac}, and retaining only $x^2$ terms, one finds the current variation $\Delta \bar{I}=(ea/(2\hbar\omega))^2[I(V+\hbar\omega/e)-2I(V)+I(V-\hbar\omega/e)]$. This is physically equivalent to retaining only the contribution of the two first sidebands with $n=\pm 1$~\cite{Tucker:1985}. This simple approximation already shows that $P_{\mathrm{int}}$ is indeed reduced by increasing the amplitude $a$. Furthermore, when the amplitude becomes comparable to the matching peak voltage $V_p$,
%or with the voltage difference $V_S-V_p$
the AC signal starts to explore points of the $I(V)$ characteristic curve with positive current ($I>0$) and the averaged thermocurrent needs to be correspondingly suppressed.

The previous discussion shows that the PAT in general just redistributes the thermoelectric power but does not increase, on average, the intrinsic thermoelectric capabilities of the junction. This is not unexpected, since also for the cooling properties of $N$-$I$-$S$ junction a similar result is observed for the cooling capabilities in presence of PAT~\cite{Kopnin:2008,Muhonen:2012,Tan:2017,Silveri:2017}. However, the redistribution due to PAT determines new values of voltages (sidebands) where the system is strongly thermoelectric for biases where the thermoelectric performance was originally smaller. For such  particular values of the bias  one observes a microwave \emph{enhanced} bipolar thermoelectricity.  Furthermore, in the absence of an AC source, the thermoelectric peaks can be very narrow in voltage range, so a change in the operating point (DC voltage) could result in a drastic reduction of the thermoelectric power. This may happen, for example, with a change in the load resistance when the junction operates as a thermoelectric generator. The AC source widening the $\bar{I}(V)$ characteristic thermoelectric curve, reduces the issue of a precise biasing. At the same time, it allows to tune the voltages where the thermoelectric effect is maximized by varying $\omega$, and in this way it  increases the tunability of the device. In other words, there are regimes in the biases where the thermoelectricity is literally microwave-assisted showing a unique interplay between thermoelectricity and coherent absorption/emission of photons. This increased tunability and better performance at specific biases may be relevant for some specific applications. Furthermore, we expect that the photon detection in $S$-$I$-$S'$ junctions can be quantum limited~\cite{Paolucci:2023-arXiv,Germanese:2023-Patent} analogously to what has been reported for tunnel junctions in the equilibrium (dissipative) case~\cite{Tucker:1979,Tucker:1985}.

In Fig.~\ref{Fig:power} we analyze the power generated by the thermoelectric effect. Figure~\ref{Fig:power}(a) shows the maximum power for a given $a$ and $\omega$, i.e. the point of the $\bar{I}$ characteristic curve which maximizes the thermoelectric power $P=\operatorname{max}_V(-\bar{I}(V)V)$. For low values of $a$, the maximum power still coincides with the central peak. But increasing the amplitude of the AC source decreases the depth of the central peak, reducing the maximum power accordingly. Above a certain amplitude, roughly delimited by the lower dashed line in the plot, the maximum power shows an oscillatory behaviour. This dashed line represents the points where the weight of the first terms in the series given by Eq. \eqref{Eq:current_ac} becomes greater than the zeroth term with increasing amplitude, i.e. $J_1(ea/\hbar\omega)=J_0(ea/\hbar\omega)$ for $ea/\hbar\omega=1.43$. Therefore, above this line, the maximum power can also be found in one of the sidebands. In the presence of PAT, the power of the sidebands varies significantly with $\omega$, generating this complex oscillatory behaviour presented in the figure. The middle ($ea/\hbar\omega=2.63$) and upper ($ea/\hbar\omega=3.77$) dashed lines correspond to the boundaries where the second ($|n|=2$) and third ($|n|=3$) terms become predominant in Eq. \eqref{Eq:current_ac}, respectively. As shown in the plot, there is a shift of the oscillating pattern delimited by the middle dashed line, and the oscillating behaviour almost vanishes above the upper line.

%
%
%%%%%%%%%%%%
\begin{figure*}[!t]
\centering
  \includegraphics[width=0.8\textwidth]{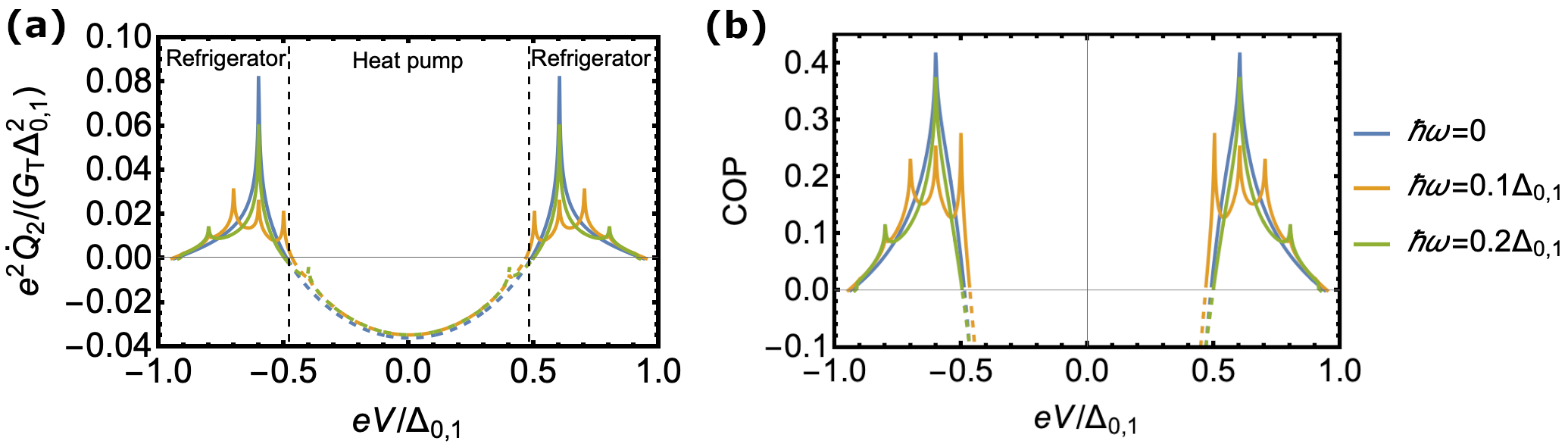}
\caption{(a) Heat current (cooling power) extracted from the right (cold) superconducting electrode in the cooling regime and (b) coefficient of performance (COP) for different frequencies. The amplitude of the AC voltage is set to $ea=0.15\Delta_{0,1}$. The temperatures of the superconducting electrodes are $T_1=0.5T_{c,1}$ and $T_2=0.4T_{c,1}$.}\label{Fig:cooling}
\end{figure*}
%%%%%%%%%%%
%
%
In Fig.~\ref{Fig:power}(b) we show the Seebeck voltage $V_S$ dependence on the amplitude $a$ taken at different frequencies $\omega$. $V_S$ increases in steps by increasing the amplitude. As discussed above, the influence of $a$ on the thermoelectric effect is to redistribute the power to other voltages by increasing the number of the thermoelectric sideband peaks. Each step is associated to the contribution of additional sidebands that once they cross the $\bar{I}=0$ line shift the Seebeck voltage sharply. However, beyond a certain value of $a$, as discussed above, the suppression of the thermoelectric effect leads to a drop of the Seebeck voltage, as shown by the $\hbar\omega=0.06\Delta_{0,1}$ (blue) line.

The $S$-$I$-$S'$ junction can be used as a thermoelectric power source by replacing the  external DC voltage source with a resistive load $R$. Ohm's law constrains the values that the current and the voltage can take, so that the current can only flow when the junction operates as a thermoelectric generator. The operating points are determined by the intersection of the $\bar{I}(V)$ characteristic curve with a line of slope $-1/R$. Similarly to the purely DC case, only the solutions with a \textit{positive} $\mathrm{d}\bar{I}/\mathrm{d}V$ are stable in the presence of PAT~\cite{Marchegiani:2020a,Marchegiani:2020b,Germanese:2022}. As shown in Fig.~\ref{Fig:power}(c), the inclusion of an AC voltage source allows also for multiple stable working points which appear at different voltages (dashed lines) and consequently different powers (solid lines). In the example, we consider a load value of $R=6 G_T^{-1}$.  For certain AC amplitudes we get multiple stable points labelled with different $n$. The $n=0$ (blue) line corresponds to the solution associated to the matching peak, while the $n=1,2,3$ lines correspond to sidebands with a lower voltage. It is possible to find stable points with a higher voltage by choosing a higher load resistance (not shown). For $0.06 < ea/\Delta_{0,1} < 0.15$ one can find two stable working points for $n=0$ (blue), due to the main peak, and $n=1$ (yellow), corresponding to the first sideband. The system can be driven to the stable working points by applying current pulses similarly to what has been realized in Ref.~\cite{Germanese:2022} in the absence of any AC signal. Therefore, PAT can be used to design devices with an increased number of states occurring at different voltages. This, for example, would  increase the number of logic states of the proposed thermoelectric volatile memory~\cite{thermoelectric_memory_patent,Marchegiani:2020d,Guarcello:2018}.

Figure~\ref{Fig:power}(d) shows the thermal efficiency $\eta=\dot{W}/\dot{Q}_1$, which describes the ratio of the net work output to the heat input, for different values of $a$. As in the case of the power, the inclusion of an AC source does not offer a way to increase globally the maximum efficiency. However, the widening of the curves is beneficial to avoid efficiency drops stemming from changes in the voltage operating point. The dashed vertical lines in Fig.~\ref{Fig:power}(d) show the area where the system is thermoelectric $V<V_S$, and where it is dissipative in the absence of the AC signal. We see that PAT \textit{widens} the voltage values where the system is thermoelectric. The AC signal leads to a moderate efficiency reduction at the matching peak, but it increases its efficiency for certain values of $V$. 
%Given the nonlinearity of the problem, it is difficult to give a general statement about the role of microwave-assisted physics over thermodynamic thermoelectric efficiency. At very low amplitudes we expect that the two first sidebands emerge from the background thermoelectric behaviour (blue line). For the $n=-1$ sideband, certainly, the thermocurrent is sustained also by the absorption of a microwave photon. So, for a such specific case, one can speculate that the photon absorption helps the thermocurrent and increases the engine efficiency. Anyway, a similar effect happens also for the photon emission sidebands $n=1$ showing that there is no specificity between emission and absorption photon processes in the presented system. This is due to the fact that the microwave contribution is computed by adiabatic approximation which corresponds to purely classical driving. 

Finally, in Fig.~\ref{Fig:cooling} we investigate the cooling power of the $S$-$I$-$S'$ junction. 
%Refrigeration corresponds to a heat engine that works in the backwards direction. Instead of extracting power from a system with a temperature difference, a refrigerator causes heat to flow from the cold terminal to the hot terminal by inputting work.
By applying an external bias, for specific temperature conditions~\cite{Marchegiani:2020b}, it is possible to extract heat and reduce the electronic temperature of the lower gap superconductor~\cite{Giazotto:2006,Muhonen:2012,Marchegiani:2020b}.
%Thermoelectricity and refrigeration are complementary effects in the linear regime. However, the second law of thermodynamics prevents a thermodynamical machine from operating as a heat engine and a cooler at the same time.
In Fig.~\ref{Fig:cooling}(a) we show the heat extracted from the lower gap superconductor for temperatures $T_1=0.5T_{c,1}$ and $T_2=0.4T_{c,1}$. The voltage range where there is no cooling is represented by dashed lines. If the applied voltage is small, the system does not reach the cooling regime and the heat flows from the hot to the cold terminal. The cooling shares some similarities with the thermoelectric effect; it requires the hot superconductor to possess a larger gap than the cold superconductor and the cooling power is maximized at voltage bias $eV=\Delta_1-\Delta_2$. With PAT the AC voltage source creates additional sidebands to the cooler, \textit{widening} the window where the cooling is possible~\cite{Kopnin:2008}. Figure~\ref{Fig:cooling}(b) shows the coefficient of performance $\mathrm{COP}=-\dot{Q}_2/\dot{W}$ which describes the efficiency of the extracted heat with respect to the applied work. Similarly to the thermoelectric efficiency, the maximum $\mathrm{COP}$ is slightly suppressed by the presence of PAT since also in this case the PAT will not increase globally the efficiency of the system but redistribute the cooling capabilities over different cooling sidebands. Analogously to the thermoelectricity there are specific values of the bias where the cooling is microwave-enhanced.

\section{Conclusions}\label{conclusions}
In this work, we have discussed the influence of photon-assisted tunneling on the bipolar thermoelectric effect occurring in an $S$-$I$-$S'$ tunnel junction. The AC voltage source leads to a weighted replication of the bipolar thermoelectric DC characteristic curve displaced in voltages. This  leads to the appearance of sidebands whose position is determined by the frequency of the AC field. The redistribution of the power to other voltages leads globally to a reduction of the net thermopower output of the matching peak, but it broadens the thermoelectric region potentially increasing even the obtainable Seebeck voltage. Therefore, changes in the operating point of the junction will have a less dramatic effect over the thermoelectric performance. We have specifically investigated how this redistribution mechanism of the bipolar thermoelectricity is influenced by the AC signal amplitude. The photon emission/absorption sidebands for the bipolar thermoelectricity constitute, in a certain sense a \emph{microwave-assisted thermoelectricity}. The thermoelectric power and the efficiency for some specific values of the voltages is even enhanced.
%Anyway, in our computation, there is not quantum specificity to differentiate between emission/absorption processes in full agreement with the adopted PAT adiabatic (semiclassical) approach which is expected to be valid in the tunneling (lowest order) limit considered here.

Furthermore, the system can be driven to the additional stable working points selected by the sidebands, so PAT can be used to design devices with an increased number of states at different voltages, potentially increasing the number of logic states of a thermoelectric volatile memory~\cite{thermoelectric_memory_patent,Marchegiani:2020d,Guarcello:2018}. 

Finally, we have studied the influence of PAT on the cooling performance of the junction. Similar to the thermoelectric effect, the cooling power is also influenced by the presence of PAT, and the redistribution of the cooling capabilities into sidebands allows also to increase the COP at specific voltages.

From the fundamental point of view this application shows, the intriguing interplay between thermoelectricity and coherent photon absorption/emission in an experimentally accessible setup. We think that the obtained results could be relevant for quantum-limited microwave detection~\cite{Germanese:2022,Germanese:2022b}, quantum sensing and microwave-superconducting technologies.

\begin{acknowledgments}
A.H. acknowledges funding from the University of the Basque Country (Project PIF20/05). F.S.B. acknowledges  financial support from Spanish AEI through projects PID2020-114252GB-I00 (SPIRIT) and TED2021-130292B-C42, the Basque Government through grant IT-1591-22, and the A. v. Humboldt Foundation. F.S.B. and F.G. acknowledge the EU Horizon 2020 Research and Innovation Framework Programme under Grant No. 800923 (SUPERTED). F.G. and A.B. acknowledge EU’s Horizon 2020 Research and Innovation Framework Programme under Grant No. 964398 (SUPERGATE) and No. 101057977 (SPECTRUM). A.B. acknowledges the Royal Society through the International Exchanges between the UK and Italy (Grants No. IEC R2 192166 and IEC R2 212041).
\end{acknowledgments}

\nocite{*}

%\bibliography{biblio}\label{LastBibItem}% Produces the bibliography via BibTeX.

\end{document}